\documentclass[conference]{IEEEtran}

\usepackage{amsmath,amssymb}
\usepackage{cite}
\usepackage{algorithm}
\usepackage{algorithmic}
\usepackage{bm}
\usepackage{graphicx}
\usepackage{color}
\usepackage{booktabs}
\usepackage{hyperref}
\usepackage{balance}

\definecolor{revise}{RGB}{0, 0, 0}

\title{Game-Theoretic Multi-Agent Reinforcement Learning for Swarm Trajectory Planning in Low-Altitude Wireless Networks \vspace{-5pt}}

\author{
\IEEEauthorblockN{Nguyen Duc Minh Quang\IEEEauthorrefmark{1}, 
Ruoxi Chong\IEEEauthorrefmark{2},
Zhiqiang Wei\IEEEauthorrefmark{3},
Chang Liu\IEEEauthorrefmark{1},
and Derrick Wing Kwan Ng\IEEEauthorrefmark{4}}
\IEEEauthorblockA{\IEEEauthorrefmark{1}School of Computing, Engineering, and Mathematical Sciences, La Trobe University, Australia}
\IEEEauthorblockA{\IEEEauthorrefmark{2}Centre for Wireless Innovation, Queen’s University Belfast, U.K.}
\IEEEauthorblockA{\IEEEauthorrefmark{3}School of Mathematics and Statistics, Xi'an Jiaotong University, China}
\IEEEauthorblockA{\IEEEauthorrefmark{4}School of Electrical Engineering and Telecommunications, University of New South Wales, Australia}
Email: \{quang.nguyen, c.liu6\}@latrobe.edu.au, rchong02@qub.ac.uk, zhiqiang.wei@xjtu.edu.cn, w.k.ng@unsw.edu.au
\vspace{-10pt}
}

\begin{document}
\maketitle


\begin{abstract}
The Low-Altitude Economy (LAE) is rapidly expanding, giving rise to low-altitude wireless networks (LAWNs), where large-scale cellular-connected unmanned aerial vehicle (UAV) deployments support heterogeneous mission-critical applications over multi-cell ground base station (GBS) infrastructures. To ensure mission success, each UAV must jointly optimize communication throughput and mission completion efficiency. In fifth-generation (5G) new radio (NR) systems, the equal resource block (RB) allocation policy induces strong strategic coupling among UAV trajectories: when a UAV enters a GBS cell, it reduces the RB share available to all co-served UAVs, thereby altering their achievable rates and trajectory incentives through shared wireless resources. Existing studies either ignore this coupling or focus on single-cell infrastructure, leaving the multi-cell, congestion-aware UAV trajectory planning problem insufficiently addressed. To fill this gap, we formulate the problem as a cooperative stochastic congestion game with a communication-and-mission-aware utility function, and propose a centralized-training decentralized-execution multi-agent proximal policy optimization (CTDE-MAPPO) algorithm to maximize social welfare under multi-cell RB congestion. Simulation results show that the proposed method outperforms QMIX, independent Q-learning, and random baselines in terms of aggregate utility and mission success rate, while achieving stable convergence within practical training budgets.  
\end{abstract}


\section{Introduction}
\label{sec:intro}

The Low-Altitude Economy (LAE) is rapidly growing, with cellular-connected unmanned aerial vehicles (UAVs) increasingly deployed for diverse mission-critical applications \cite{Zeng2019, nguyen2024deep, Yang2026}. This evolution is giving rise to low-altitude wireless networks (LAWNs), where multiple UAVs operate over shared multi-cell ground base station (GBS) infrastructures~\cite{liu2026scalablepredictive}. Unlike conventional single-UAV scenarios, UAV trajectory decisions in LAWNs are inherently coupled through shared wireless resources and inter-cell interactions~\cite{liu2022learning, liu2020deepresidual}. Under the equal resource block (RB) allocation policy widely adopted in fifth-generation (5G) new radio (NR) networks \cite{Zhang2021RadioMap}, a UAV entering a congested GBS cell reduces the RB share of all co-served UAVs, thereby degrading their achievable rates. Meanwhile, each UAV must complete its mission efficiently \cite{Yang2026}. Thus, each UAV’s optimal trajectory depends on not only its own destination and channel condition, but also the real-time GBS associations of other UAVs. Since centralized coordination is impractical due to communication overhead, computational complexity, and limited scalability \cite{Wu2020TCOM}, a decentralized trajectory planning framework is needed to capture multi-cell shared-RB coupling while balancing congestion avoidance and mission completion efficiency.

UAV trajectory optimization for communication has been extensively studied, and multi-agent reinforcement learning (MARL)~\cite{albrecht2024multi} has emerged as a promising approach for sequential decision-making in non-stationary environments with decentralized execution. Existing MARL-based studies, such as \cite{Wu2020TCOM}, apply multi-agent learning to UAV control in cellular networks but treat each UAV's wireless link independently, thereby neglecting RB-sharing effects among co-served UAVs. Other joint trajectory and resource allocation work \cite{Ye2025ISAC} considers multi-agent settings but focus mainly on single-cell infrastructure, where UAV mobility across multiple GBSs is not modeled. Consequently, the joint optimization of communication throughput and mission completion time for multiple mission-driven UAVs competing for shared RBs across multi-GBS networks remains largely unexplored.

Given the strategic coupling nature of this problem, \emph{game theory}, with its ability to formally model rational agents with interdependent payoffs~\cite{Yang2026}, provides a principled tool to formulate such interactions. This paper addresses this gap through the following contributions:
\begin{itemize}
  \item Unlike existing works that either ignore RB sharing effects or assume single-cell infrastructure, we investigate the multi-cell LAWN setting under equal RB allocation, formulating the multi-UAV trajectory planning problem as a cooperative stochastic congestion game.
  \item We model the congestion game as an MDP and propose CTDE-MAPPO with a congestion-aware observation space using GBS cell loads as the coupling sufficient statistic, and a per-step reward decomposition that approximates the game utility throughout training.
  \item Simulation results demonstrate superior aggregate utility and mission success rate over QMIX, independent Q-learning, and random baselines, with stable convergence within practical training budgets.
\end{itemize}

\section{System Model}
\label{sec:system}

\begin{figure}[t]
  \centering
  \includegraphics[width=3.5 in]{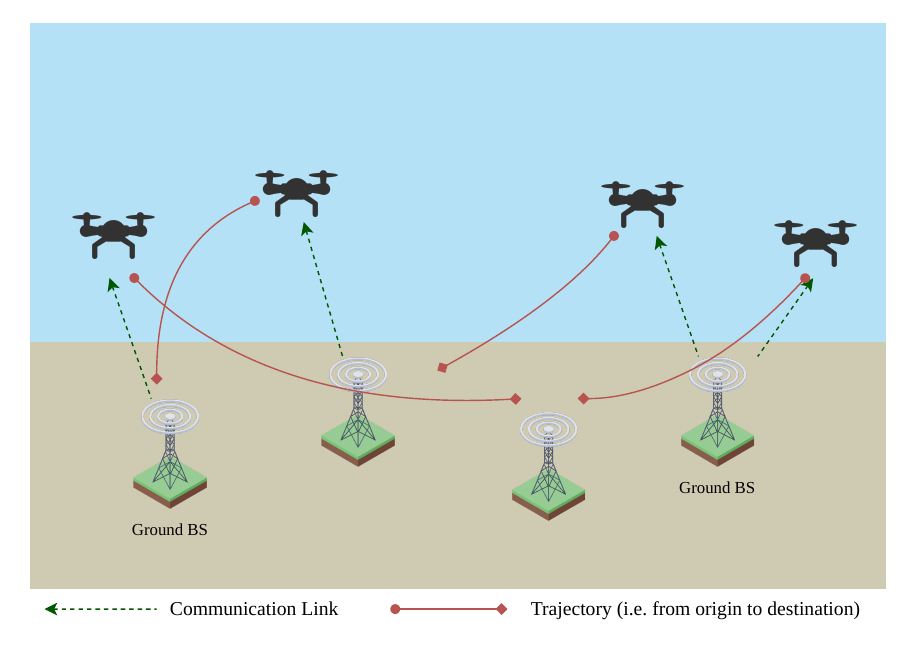}
  \caption{Illustration of the multiple cellular-connected UAV working in LAWN.}
  \label{fig:system}
\end{figure}

We consider a cellular-connected low-altitude network consisting of $M$ GBSs and $N$ UAVs, as illustrated in Fig.~\ref{fig:system}. Let $\mathbf{q}_i(t) \in \mathbb{R}^3$ denote the 3D position of UAV $i \in \mathcal{N} = \{1,\ldots,N\}$ at time $t$. Each GBS $b \in \mathcal{B} = \{1,\ldots,M\}$ is located at a fixed known position $\mathbf{g}_b$ and serves a subset of UAVs within its coverage area using a total of $K_b$ resource blocks (RBs), each of sub-bandwidth $\Delta f$ Hz.

\subsection{Communication Model}

Each GBS serves the UAVs associated with it over shared RBs. Let $\mathcal{U}_b(t) = \{i : b_i(t) = b\}$ denote the set of UAVs associated with GBS $b$ at time $t$, where $b_i(t) = \arg\min_{k \in \mathcal{B}} \|\mathbf{q}_i(t) - \mathbf{g}_k\|$ is the nearest-GBS association rule, and let $n_b(t) = |\mathcal{U}_b(t)|$ denote the instantaneous cell load. Under the equal RB allocation policy, which is widely adopted in 5G NR systems \cite{Zhang2021RadioMap}, GBS $b$ distributes its $K_b$ RBs equally among all $n_b(t)$ currently associated UAVs. The number of RBs allocated to UAV $i \in \mathcal{U}_b(t)$ is therefore
\begin{equation}
  \kappa_{b,i}(t) = \left\lfloor \frac{K_b}{n_b(t)} \right\rfloor.
  \label{eq:rb}
\end{equation}
Under the LoS-dominant air-to-ground propagation model\footnote{\textcolor{revise}{This work assumes interference-free transmission, where the RB assigned to each UAV is not reused by neighboring GBSs. Inter-cell interference is left for future work.}}, which is appropriate for UAVs operating above obstacle clutter height \cite{Al-Hourani2014,liu2019deep}, the achievable rate of UAV $i$ at time $t$ is given by
\begin{equation}
  R_i(t) = \kappa_{b_i,i}(t) \cdot \Delta f \log_2\!\left(1 + \frac{P_{\mathrm{tx}}\,\beta_0}{\sigma^2 \|\mathbf{q}_i(t) - \mathbf{g}_{b_i(t)}\|^{\alpha}}\right),
  \label{eq:rate}
\end{equation}
where $P_{\mathrm{tx}}$ is the fixed per-RB transmit power, $\beta_0$ is the reference channel gain at unit distance, $\alpha$ is the path-loss exponent, $\sigma^2$ is the noise power per RB, and $\|\mathbf{q}_i(t) - \mathbf{g}_{b_i(t)}\|$ denotes the physical distance. 

\subsection{UAV Mobility Model}

The operational airspace is bounded within the altitude range $[h_{\min}, h_{\max}]$. Following the spatial discretization framework in \cite{quang20253d}, the 3D airspace is partitioned into a uniform voxel grid $\mathcal{V} \in \mathbb{R}^{W \times L \times H}$ with spatial resolution $\Delta_s$. Within each voxel, large-scale propagation conditions are assumed approximately constant, so that the channel gain in Eq.~\eqref{eq:rate} can be evaluated at the voxel center without loss of generality. Time is discretized with slot duration $\delta_t$, chosen such that UAV displacement within a single slot remains within one voxel. All positions $\mathbf{q}_i(t)$ are defined at voxel centers throughout this paper.

At each time slot $t$, UAV $i$ is located at voxel $\mathbf{q}_i(t) \in \mathcal{V} \subset \mathbb{Z}^3$ and selects a movement action $\mathbf{a}_i(t) \in \mathbb{Z}^3$, representing the displacement in the three spatial dimensions. The position evolves as
\begin{equation}
  \mathbf{q}_i(t+1) = \mathbf{q}_i(t) + \mathbf{a}_i(t),
  \label{eq:mobility}
\end{equation}
where $v_{\max} \in \mathbb{N}$ is the maximum number of voxel steps per slot, $\|\mathbf{a}_i(t)\|_1 \leq v_{\max}$, and $\mathbf{q}_i(t+1) \in \mathcal{V}$. To prevent physical conflicts among UAVs sharing the same airspace, the following collision avoidance constraint must be satisfied at all times:
\begin{equation}
  \mathbf{q}_i(t) \neq \mathbf{q}_j(t), \quad \forall\, i \neq j \in \mathcal{N},\; \forall\, t.
  \label{eq:collision}
\end{equation}
Mission $i$ is declared complete at the first slot $T_i$ when UAV $i$ reaches the destination voxel containing $\mathbf{d}_i$, subject to the deadline constraint:
\begin{equation}
  T_i \leq T_{\max}, \quad \forall\, i \in \mathcal{N}.
  \label{eq:deadline}
\end{equation}
The trajectory of UAV $i$ is defined as the sequence of visited voxels from origin to destination, $\mathcal{S}_i = \{\mathbf{q}_i(t)\}_{t=0}^{T_i}$, and the joint trajectory of all UAVs is denoted $\mathcal{S} = \{\mathcal{S}_i\}_{i \in \mathcal{N}}$.

\section{Game-Theoretic Problem Formulation}
\label{sec:formulation}

The RB allocation rule in Eq.~\eqref{eq:rb} naturally induces a congestion mechanism. Specifically, as more UAVs associate with the same GBS, the RB share allocated to each co-served UAV becomes non-increasing, leading to the coupling
$n_b(t) \uparrow \Rightarrow \kappa_{b,i}(t) \downarrow \Rightarrow R_i(t) \downarrow$
for a fixed link geometry. Since $n_b(t)$ is determined by the positions of all UAVs, each UAV's communication rate is affected by the trajectory decisions of the others. This creates strategic coupling that cannot be captured by independent single-agent trajectory optimization. We therefore model the problem as a cooperative stochastic congestion game \cite{Rosenthal1973}:
\begin{equation}
  \mathcal{G} = \langle \mathcal{N}, \mathcal{S}, \{\mathcal{A}_i\}_{i \in \mathcal{N}}, P, \{U_i\}_{i \in \mathcal{N}} \rangle,
\end{equation}
where $\mathcal{N}$ is the set of UAV players, $\mathcal{S}$ denotes the joint trajectory space, $\mathcal{A}_i$ is the strategy space of UAV $i$, $P$ characterizes the state transition induced by UAV mobility, and $U_i$ is the payoff function of UAV $i$. This game-theoretic formulation enables the congestion externality caused by shared RB allocation to be explicitly captured, while allowing the collective objective to be expressed through social welfare maximization. It also motivates a MARL-based solution that exploits the underlying game structure.

\subsection{Player Utility and Social Welfare}

Each UAV seeks to maximize communication throughput while completing its mission efficiently. However, the sum rate $\sum_t R_i(t)$ is an inadequate metric as it grows with mission duration and fails to penalize rate fluctuation caused by GBS congestion. Thus, we instead adopt the exponential moving average (EMA) \cite{Polyak1992} of the rate sequence:
\begin{equation}
  \bar{R}_i(t) = (1-\beta)\,\bar{R}_i(t-1) + \beta R_i(t), \quad \bar{R}_i(0) = 0.
  \label{eq:ema}
\end{equation}
Here $\bar{R}_i(t)$ is a recursive accumulation of the rate history $\{{R}_i(t)\}_{\tau=0}^t$ with $\bar{R}_i(0) = 0$ and smoothing factor $\beta \in (0,1)$, whose terminal value $\bar{R}_i(T_i)$ captures both mean rate quality and temporal consistency~\cite{Polyak1992}. To ensure fairness across missions of different origin-destination distances, we measure efficiency relative to the minimum flight time
\begin{equation}
  T_i^{\min} = \left\lceil \frac{\|\mathbf{q}_i^0 - \mathbf{d}_i\|_1}{v_{\max}} \right\rceil,
  \label{eq:tmin}
\end{equation}
defined as the minimum slots to complete mission $i$ at maximum speed. The ratio $T_i^{\min}/T_i \in (0,1]$ equals unity on the time-optimal path and decreases monotonically with detour overhead, providing a mission-agnostic efficiency measure. Since communication quality and mission efficiency are complementary, i.e., a UAV achieving an excellent rate but never arriving, or arriving quickly through poor-coverage cells, both constitute mission failures, we define the player payoff as their product \cite{Nash1950}:
\begin{equation}
  U_i = \hat{R}_i(T_i) \cdot \frac{T_i^{\min}}{T_i},
  \label{eq:utility}
\end{equation}
where $\hat{R}_i(T_i) = \bar{R}(T_i)/(1-(1-\beta)^{\bar{T}})$ denotes the bias-corrected terminal EMA rate, which is used to mitigate initialization bias for short missions. The main objective of this study is to maximize the social welfare, defined as the sum of individual utilities over all UAVs:
\begin{equation}
  \max_{\mathcal{S}}\; W(\mathcal{S}) = \sum_{i \in \mathcal{N}} U_i, \quad \text{s.t.}~\eqref{eq:mobility}, \eqref{eq:collision}, \eqref{eq:deadline}.
  \label{eq:welfare}
\end{equation}

\subsection{MDP Formulation}

Since the game unfolds over discrete time slots with partial information, we operationalize $\mathcal{G}$ as an MDP defined by the tuple $\langle \mathcal{N}, \mathcal{X}, \{\mathcal{O}_i\}, \{\mathcal{A}_i\}, \{r_i\}, P, \gamma \rangle$, where $\mathcal{X}$ is the global state space, $\mathcal{O}_i$ is the local observation space, $r_i$ is the per-step reward, and $\gamma \in (0,1)$ is the discount factor used for variance reduction during training. Each UAV maintains a local policy $\pi_{\theta_i}: \mathcal{O}_i \rightarrow \Delta(\mathcal{A}_i)$, and the joint objective follows Eq.~\eqref{eq:welfare}.

\textit{Global State.} The centralized critic accesses during training:
\begin{equation}
  \mathbf{x}(t) = \left[\{\mathbf{q}_i(t), \mathbf{d}_i, T_i^{\min}, \bar{R}_i(t)\}_{i \in \mathcal{N}},\ \{n_b(t), \mathbf{g}_b\}_{b \in \mathcal{B}}\right],
  \label{eq:global_state}
\end{equation}
giving full visibility of the congestion coupling across all agents, enabling correct credit assignment for cooperative decisions invisible to any individual agent.

\textit{Local Observation.} At execution, each UAV acts on:
\begin{equation}
  \mathbf{o}_i(t) = \left[\mathbf{q}_i(t), \mathbf{d}_i, T_i^{\min}, \bar{R}_i(t),\ \{n_b(t), \mathbf{g}_b\}_{b \in \mathcal{B}}\right],
  \label{eq:obs}
\end{equation}
where $(\mathbf{q}_i, \mathbf{d}_i, T_i^{\min})$ provides navigation context, $\bar{R}_i(t)$ reflects link quality history, and $\{n_b(t)\}$ serves as the congestion sufficient statistic for anticipating rate degradation upon cell entry.

\textit{Action Space.} The per-slot action space inherits from the game strategy space as:
\begin{equation}
  \mathcal{A}_i = \left\{\mathbf{a} \in \mathbb{Z}^3 : \|\mathbf{a}\|_1 \leq v_{\max}\right\},
  \label{eq:action_space}
\end{equation}
including the null action $\mathbf{a} = \mathbf{0}$ (hover). Solving $\mathcal{G}$ directly is intractable: the joint strategy space grows exponentially with $N$, the congestion coupling introduces non-stationarity, and the terminal payoff $U_i$ cannot be decomposed into per-step decisions analytically. MARL \cite{albrecht2024multi} can address these challenges by operating without an explicit environment model, handling coupled state spaces naturally, and supporting decentralized execution.

\section{Proposed CTDE-MAPPO Solution}
\label{sec:marl}
This section presents the proposed solution to the optimization problem in Eq.~\eqref{eq:welfare}. We first design a per-step reward decomposition that provides dense learning signals while approximating the terminal game payoff in Eq.~\eqref{eq:utility}. We then describe the actor--critic network architecture and the CTDE-MAPPO training procedure.

\subsection{Reward Design}

The terminal game payoff $U_i$ in Eq.~\eqref{eq:utility} is realized only when mission $i$ is completed, which leads to sparse feedback over long planning horizons. We therefore design a per-step reward $r_i(t)$ for the transition from $t$ to $t+1$ to provide dense learning signals for both communication quality and navigation, while retaining the terminal payoff as the final objective:
\begin{equation}
  r_i(t) = r_i^{\mathrm{rate}}(t) + r_i^{\mathrm{nav}}(t) + r_i^{\mathrm{col}}(t) + r_i^{\mathrm{term}}(t).
  \label{eq:reward}
\end{equation}

The rate-quality signal is defined as the incremental change in the EMA rate:
\begin{equation}
  r_i^{\mathrm{rate}}(t) = \bar{R}_i(t+1) - \bar{R}_i(t),
  \label{eq:reward_rate}
\end{equation}
which provides positive feedback when the updated rate history improves and negative feedback when it deteriorates. Since this term telescopes over time, it promotes communication quality without directly favoring longer trajectories.

The navigation shaping signal encourages progress toward the destination:
\begin{equation}
  r_i^{\mathrm{nav}}(t)
  =
  \omega
  \left(
  \|\mathbf{q}_i(t)-\mathbf{d}_i\|
  -
  \|\mathbf{q}_i(t+1)-\mathbf{d}_i\|
  \right),
  \label{eq:reward_nav}
\end{equation}
where $\omega>0$ is the shaping coefficient. This term provides positive feedback when UAV $i$ moves closer to its destination, negative feedback when it moves farther away, and zero feedback when the remaining distance is unchanged.

The collision penalty discourages violations of the collision avoidance constraint in Eq.~\eqref{eq:collision} during training:
\begin{equation}
  r_i^{\mathrm{col}}(t) = -\rho_{\mathrm{col}} \mathbf{1}\!\left[\exists\, j \neq i : \mathbf{q}_j(t+1) = \mathbf{q}_i(t+1)\right],
  \label{eq:reward_col}
\end{equation}
where $\rho_{\mathrm{col}} > 0$ is the weight of the collision penalty.

The terminal reward anchors the learning objective to the game payoff:
\begin{equation}
  r_i^{\mathrm{term}}(t) =
  \begin{cases}
    U_i, & \text{if } \mathbf{q}_i(t+1) = \mathbf{d}_i,\ T_i \leq T_{\max}, \\[4pt]
    -r_{\mathrm{fail}}, & \text{if } t+1 = T_{\max},\ \mathbf{q}_i(T_{\max}) \neq \mathbf{d}_i, \\[4pt]
    0, & \text{otherwise,}
  \end{cases}
  \label{eq:terminal}
\end{equation}
where $r_{\mathrm{fail}} > 0$ penalizes mission timeout. The dense rate and navigation terms guide the learning process throughout the episode, while the terminal term aligns the learned policy with the payoff in Eq.~\eqref{eq:utility}.

\subsection{Network Architecture and Training}

\begin{algorithm}[t]
\caption{CTDE-MAPPO for Multi-UAV Trajectory Planning}
\label{alg:mappo}
\begin{algorithmic}[1]
\REQUIRE Actor $\theta$, critic $\phi$, clip ratio $\epsilon$, entropy coefficient $c_e$, GAE parameter $\lambda_{\text{GAE}}$, rollout length $L$, update epochs $K$, total iterations $I$
\FOR{iteration $= 1, 2, \ldots, I$}
  \STATE Reset environment; initialize UAVs at $\{\mathbf{q}_i^0\}$; compute $\{T_i^{\min}\}$; set $\bar{R}_i(0) = 0,\ \forall i$
  \FOR{step $t = 0, 1, \ldots, L-1$}
    \STATE Each UAV $i$ observes $\mathbf{o}_i(t)$; sample $\mathbf{a}_i(t) \sim \pi_\theta(\cdot|\mathbf{o}_i(t))$
    \STATE Execute $\{\mathbf{a}_i(t)\}$; collect $\{r_i(t)\}$ and $\mathbf{x}(t+1)$; update $\bar{R}_i(t)$ via ~\eqref{eq:ema}
    \STATE Reset UAV $i$ to new mission if $\mathbf{q}_i(t+1) = \mathbf{d}_i$ or $t+1 = T_{\max}$
  \ENDFOR
  \STATE Compute TD errors $\{\delta_t^i\}$ and GAE advantages $\{\hat{A}_t^i\}$ via \eqref{eq:gae} using $V_\phi(\mathbf{x}(t))$
  \STATE Compute returns $\hat{G}_t^i = \hat{A}_t^i + V_\phi(\mathbf{x}(t))$; normalize $\hat{G}_t^i$ to zero mean and unit variance
  \FOR{epoch $k = 1, \ldots, K$}
    \STATE Update $\theta$ by maximizing $\mathcal{L}^{\text{clip}}(\theta) + c_e H(\pi_\theta)$ via \eqref{eq:ppo_clip}
    \STATE Update $\phi$ by minimizing $\mathcal{L}^{\text{critic}}(\phi)$ via \eqref{eq:critic_loss}
  \ENDFOR
\ENDFOR
\STATE \textbf{Execution:} deploy actor $\pi_\theta$ only; each UAV acts on local $\mathbf{o}_i(t)$, no critic or communication needed
\end{algorithmic}
\end{algorithm}

CTDE-MAPPO is selected because the non-linear RB congestion coupling makes the joint value function non-decomposable, ruling out value decomposition methods such as QMIX~\cite{rashid2020monotonic}, while the continuous trajectory action space favors policy gradient over Q-learning~\cite{tan1993multi}. Critically, the centralized critic observes the full congestion state $\{n_b(t)\}$ during training, enabling correct credit assignment when one UAV's rate is directly affected by others' cell occupancy. At execution, each decentralized actor conditions only on its local observation $\mathbf{o}_i(t)$, satisfying the practical constraint of no inter-UAV communication~\cite{liu2020deeptransfer}. The network details, training, and execution procedure are presented as follows.

\textit{Actor Network.} Each UAV adopts a shared decentralized actor
$\pi_{\theta}: \mathcal{O}_i \rightarrow \Delta(\mathcal{A}_i)$,
which maps the local observation $\mathbf{o}_i(t)$ to a probability distribution over the discrete action space $\mathcal{A}_i$. The actor is implemented as an MLP with two hidden layers of $256$ units and ReLU activations, followed by a softmax output layer. The sharing of parameters among UAVs exploits the homogeneity of the agent, reduces the number of trainable parameters, and improves the efficiency of the sample in cooperative MARL~\cite{Yu2022MAPPO}.

\textit{Critic Network.} A single centralized critic $V_{\phi}: \mathcal{X} \rightarrow \mathbb{R}$ maps the global state $\mathbf{x}(t)$ to a scalar estimate of the expected discounted team return. The critic is implemented as an MLP with two hidden layers of 256 units and ReLU activations. To improve training stability under the variable-magnitude rewards induced by the product-form utility in \eqref{eq:utility}, value normalization is applied to the return targets during critic training \cite{Yu2022MAPPO}.

\textit{Training and Execution Procedure.} Training proceeds in iterations, each consisting of two phases. In the collection phase, all $N$ UAVs interact with the environment for $L$ steps following the current actor policy $\pi_\theta$, storing observations, actions, rewards, and global states in a shared rollout buffer. In the update phase, both networks are trained on this buffer before fresh experience is collected.

\textcolor{revise}{The \textit{advantage} $\hat{A}_t^i$ measures how much better the action taken by UAV $i$ at step $t$ was relative to the average action under the current policy, and is estimated using Generalized Advantage Estimation (GAE) \cite{schulman2015high}:
\begin{align}
  \hat{A}_t^i &= \sum_{k=0}^{L-t} (\gamma \lambda_{\text{GAE}})^k \delta_{t+k}^i, \\
  \delta_t^i &= r_i(t) + \gamma V_\phi(\mathbf{x}(t+1)) - V_\phi(\mathbf{x}(t)),
  \label{eq:gae}
\end{align}
where $\delta_t^i$ is the temporal difference (TD) error between the reward received plus discounted future value and the current value estimate, and $\lambda_{\text{GAE}} \in [0,1]$ controls the bias-variance tradeoff.}

\textcolor{revise}{The actor is updated by maximizing the PPO clipped surrogate objective \cite{schulman2017proximal}, which prevents destructively large policy updates by clipping the probability ratio between the new and old policies:
\begin{equation}
  \mathcal{L}^{\text{clip}}(\theta) = \mathbb{E}_t\!\left[\min\!\left(\rho_t(\theta)\hat{A}_t,\ \mathrm{clip}\!\left(\rho_t(\theta), 1-\epsilon, 1+\epsilon\right)\hat{A}_t\right)\right],
  \label{eq:ppo_clip}
\end{equation}
where $\rho_t(\theta) = \pi_\theta(a_t|\mathbf{o}_t)/\pi_{\theta_{\text{old}}}(a_t|\mathbf{o}_t)$ and $\epsilon$ bounds the per-update policy change. An entropy bonus $c_e H(\pi_\theta)$ is added to encourage exploration and prevent premature convergence to a deterministic policy. The critic minimizes the mean squared error between its prediction and the empirical return:
\begin{equation}
  \mathcal{L}^{\text{critic}}(\phi) = \frac{1}{NL}\sum_{i \in \mathcal{N}}\sum_{t=0}^{L-1} \left(V_\phi(\mathbf{x}(t)) - \hat{G}_t^i\right)^2,
  \label{eq:critic_loss}
\end{equation}
where $\hat{G}_t^i = \hat{A}_t^i + V_\phi(\mathbf{x}(t))$, with value normalization applied to stabilize training under the variable-magnitude rewards of the product utility Eq.~\eqref{eq:utility}. Both networks are updated for $K$ epochs per rollout using Adam. The full procedure is summarized in Algorithm~\ref{alg:mappo}.}

At execution, the centralized critic is discarded and each UAV acts solely on its local observation Eq.~\eqref{eq:obs} with no inter-UAV communication, satisfying the decentralized execution constraint of practical LAWN deployments.

\section{Simulation Results}
\label{sec:sim}

\subsection{Simulation Setup}

\begin{table}[t]
\centering
\caption{Simulation parameters.}
\label{tab:params}
\begin{tabular}{lll}
\toprule
\textbf{Parameter} & \textbf{Symbol} & \textbf{Value} \\
\midrule
Number of UAVs             & $N$                 & $4$ \\
Number of GBSs             & $M$                 & $4$ \\
Voxel grid size            & $W \times L \times H$ & $100 \times 100 \times 30$ \\
Spatial resolution         & $\Delta_s$          & $10$ m \\
Maximum voxel steps/slot   & $v_{\max}$          & $3$ \\
Maximum episode length     & $T_{\max}$          & $100$ slots \\
Total RBs per GBS          & $K_b$               & $10$ \\
Sub-bandwidth per RB       & $\Delta f$          & $180$ kHz \\
Path-loss exponent         & $\alpha$            & $2.0$ \\
\bottomrule
\end{tabular}
\end{table}

In the simulation, the GBSs are placed at fixed ground-level locations on a regular grid, while UAV origins and destinations are randomly generated in each episode without spatial overlap. The key simulation parameters are summarized in Table~\ref{tab:params}. All results are averaged over five independent random seeds. After training, each method is evaluated over 200 test episodes without exploration.

\textbf{Baselines.} We compare the proposed CTDE-MAPPO method with three representative baselines:

\begin{itemize}
  \item \textbf{Random}: Each UAV selects an action uniformly at random from $\mathcal{A}_i$ at every time slot. This serves as a non-learning lower bound.

  \item \textbf{Independent Q-learning (IQL)}~\cite{tan1993multi}: Each UAV independently trains a Q-network using only its local observation $\mathbf{o}_i(t)$. Since no centralized critic or coordination mechanism is used, other UAVs are treated as part of a non-stationary environment.

  \item \textbf{QMIX}~\cite{rashid2020monotonic}: QMIX is a cooperative value-decomposition method that uses a monotonic mixing network to combine per-agent utilities into a joint action-value function, enabling centralized training with decentralized execution.
\end{itemize}

\subsection{Performance Comparison}

\begin{figure}[t]
  \centering
  \includegraphics[width=\columnwidth]{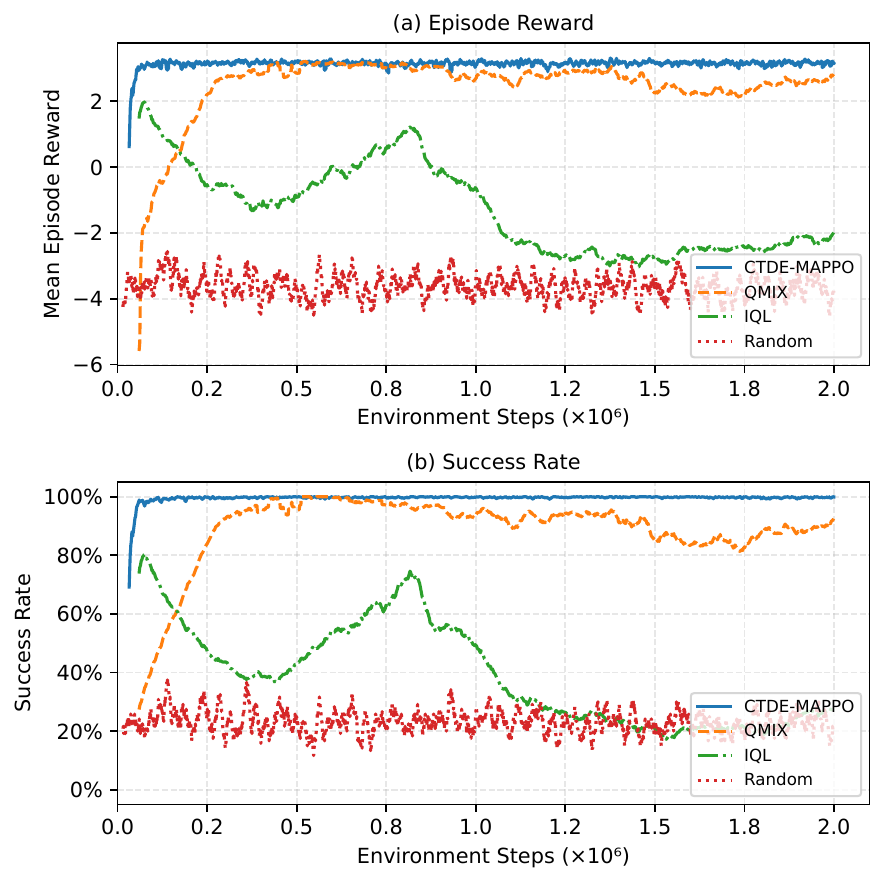}
  \caption{Training convergence: (a) mean episode reward and (b) mission
    success rate versus environment steps, smoothed with a 30-point window.}
  \label{fig:convergence}
\end{figure}

Fig.~\ref{fig:convergence} compares the training convergence of CTDE-MAPPO and the baselines in terms of mean episode reward and mission success rate, where success is defined as the percentage of UAVs reaching their destination voxels within the deadline in Eq.~\eqref{eq:deadline}. CTDE-MAPPO converges rapidly, achieving a success rate above $95$\% within approximately $150$K environment steps and remaining stable thereafter. This indicates that the centralized critic effectively leverages global congestion information during training to guide the decentralized actors toward coordinated trajectory decisions.

QMIX also converges, but with a lower and more fluctuating success rate, indicating that its monotonic value-decomposition structure is less effective in representing the nonlinear RB congestion coupling across multiple GBSs. IQL exhibits unstable learning behavior: although its success rate briefly improves at the early stage, it later degrades because each UAV treats the other learning agents as part of a non-stationary environment. The Random baseline remains almost flat, confirming that the task requires deliberate congestion-aware coordination rather than unstructured exploration~\cite{quang2026diffusion}.

\begin{table}[t]
\centering
\caption{Performance comparison (200 test episodes, mean). $\hat{R}$ is the EMA-corrected rate.}
\label{tab:perf}
\begin{tabular}{lcccc}
\toprule
\textbf{Method} & $\bar{J}$ & $\hat{R}$ (Mbps) & $\bar{T}$ (slots) & Success (\%) \\
\midrule
  Random                                   & -3.57 & 36.20 & 100.0 & 22.4\% \\
  IQL                                      & -1.96 & 44.46 & 99.7 & 25.6\% \\
  QMIX                                     & 3.01 & 38.80 & 34.6 & 96.1\% \\
  \textbf{CTDE-MAPPO}                & 3.19 & 35.68 & 8.7 & 100.0\% \\
\bottomrule
\end{tabular}
\end{table}

Table~\ref{tab:perf} reports the average per-UAV accumulated episode reward $\bar{J}$, the EMA-corrected rate $\hat{R}$, average completion time $\bar{T}$, and mission success rate. CTDE-MAPPO achieves the highest accumulated reward and perfect success rate while completing missions in significantly fewer steps. Its corrected rate $\hat{R}=35.7$~Mbps is comparable to QMIX, confirming that efficient navigation does not sacrifice communication quality. Particularly, the high $\bar{J}$ reflects simultaneous achievement of both factors of the game payoff Eq.~\eqref{eq:utility}. This stems from the centralized critic observing the full congestion state $\{n_b(t)\}$ and estimating the joint value function without structural constraints, enabling proper credit assignment when one UAV's rate is directly affected by others' cell occupancy. QMIX improves over IQL, validating cooperative value decomposition, but its lower reward reflects the expressivity limitation of the monotone mixer under asymmetric congestion interactions. IQL's higher $\hat{R}=44.5$~Mbps is a misleading artifact, i.e., agents failing to navigate tend to hover near their start positions close to lightly loaded GBSs, producing favorable channel conditions unrepresentative of deliberate optimization.

\subsection{Single-Agent Utility Distribution}

\begin{figure}[t]
\centering
\includegraphics[width=0.9\linewidth]{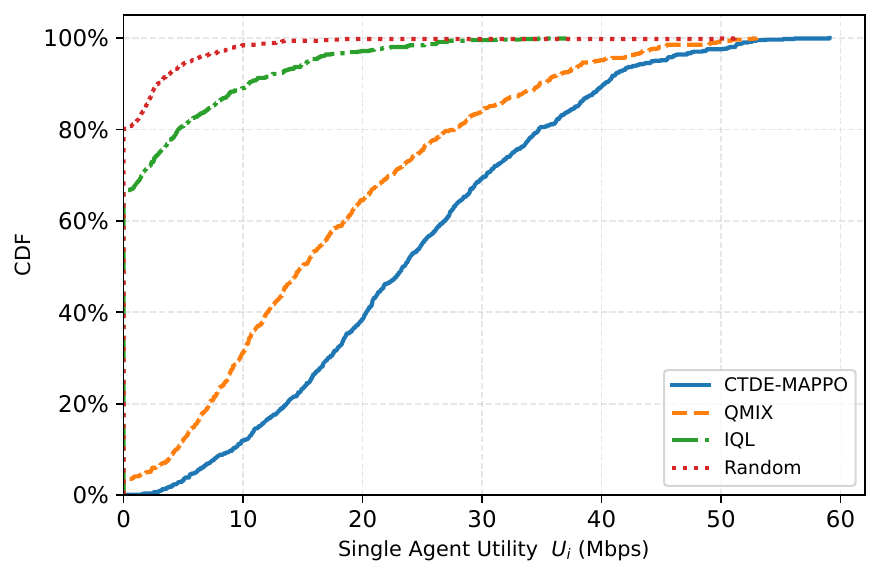}
\caption{Empirical CDF of single-agent utility $U_i$ across 200 episodes for all methods.}
\label{fig:utility_cdf}
\end{figure}

Fig.~\ref{fig:utility_cdf} presents the empirical cumulative distribution function (CDF), $F(u) = \Pr(U_i \leq u)$, of single-agent utility $U_i$ across all 200 episodes for each method. The CTDE-MAPPO curve exhibits a gradual slope spanning a wide range of $U_i$ values, revealing high variance: while many agents achieve high utility, some agents receive low utility outcomes. This indicates that the social-welfare-maximizing joint policy distributes utilities unevenly, i.e., some agents are routed through more congested cells, incurring both lower rates and longer completion times to benefit the collective. QMIX shows a more concentrated distribution, suggesting more uniform but lower utility allocation across agents. IQL and Random both collapse near zero, confirming that without cooperative congestion awareness agents fail to achieve meaningful utilities regardless of fairness. This reveals that optimizing for the social welfare does not guarantee fair outcomes for every individual agent in the congestion game, a problem that will be formally analyzed through equilibrium characterization and fairness criteria in future work.

\balance
\section{Conclusion}
\label{sec:conclusion}

This paper investigated UAV swarm trajectory planning in multi-cell LAWNs under equal RB allocation from a game-theoretic perspective. We formulated the problem as a cooperative stochastic congestion game, where shared RB allocation explicitly captures the coupling among UAV trajectories across multiple GBSs. A product-form utility was introduced to jointly characterize communication quality and mission completion efficiency, and the resulting objective was embedded into a social welfare maximization framework. To address the induced MDP, we developed a CTDE-MAPPO algorithm in which the centralized critic exploits the global congestion state during training, while decentralized actors enable practical execution without real-time centralized coordination. Simulation results demonstrated that the proposed method achieves higher aggregate utility and mission success rate than all considered baselines. The single-agent utility distribution reveals that while the social-welfare-maximizing joint policy achieves high aggregate performance, it distributes utilities unevenly between agents. Future work will extend the framework to larger-scale deployments and provide deeper game-theoretic analysis of equilibrium behavior under multi-cell congestion.
\bibliographystyle{IEEEtran}
\bibliography{refs}

\end{document}